\documentclass[12pt]{article}
\usepackage{graphicx}
\usepackage{amssymb}
\begin{document} 

\begin{titlepage}
	\rightline{}
	
	\rightline{hep-th/0608103}
	
	\vskip 2cm 
	\begin{center}
		\Large{{\bf Emergence of the fuzzy horizon\\
		through gravitational collapse}} 
	\end{center}
	
	\vskip 2cm 
	\begin{center}
		{Anand Murugan\footnote{\texttt{Anand.Murugan@pomona.edu}}\ \ \ and\ \ \ Vatche Sahakian\footnote{\texttt{sahakian@hmc.edu}}}\\
	\end{center}
	\vskip 12pt \centerline{\sl Harvey Mudd College, Pomona College} \centerline{\sl Claremont CA 91711 USA}
	
	\vskip 2cm 
	\begin{abstract}
		For a large enough Schwarzschild black hole, the horizon is a region of space where gravitational forces are weak; yet it is also a region leading to numerous puzzles connected to stringy physics. In this work, we analyze the process of gravitational collapse and black hole formation in the context of light-cone M theory. We find that, as a shell of matter contracts and is about to reveal a black hole horizon, it undergoes a thermodynamic phase transition. This involves the binding of D0 branes into D2's, and the new phase leads to large membranes of the size of the horizon. These in turn can sustain their large size through back-reaction and the dielectric Myers effect - realizing the fuzzball proposal of Mathur and the Matrix black hole of M(atrix) theory. The physics responsible for this phenomenon lies in strongly coupled $2+1$ dimensional non-commutative dynamics. The phenomenon has a universal character and appears generic. 
	\end{abstract}
\end{titlepage}

\newpage \setcounter{page}{1} 
\section{Introduction and results}

In General Relativity, the horizon of a large black hole is a cross-over line in spacetime, one that may in principle be safely crossed by an in-falling observer. While the traditional geometrical description of the gravitational dynamics is expected to break down at the center of the black hole - where curvature scales reach the Planck scale - a large enough black hole may exhibit weak gravity near its horizon. This has been the source of many of the puzzles of black hole physics.

Over the years, there have been indications that this general relativistic picture is incorrect~\cite{Horowitz:1997fr}-\cite{Mathur:2005ai}. The suggestion is that the spacetime metric of a black hole is to be trusted up to the horizon surface, beyond which a black hole appears as a fuzz with an ill-defined geometrical description. In this work, we test these ideas by considering the process of gravitational collapse and black hole formation in the context of string theory. Our goal is to zero onto the moment a horizon is to emerge and hence - if the fuzzball proposal is realized - we hope to identify the physical criterion responsible for the break down of general relativity at curvatures much smaller than the Planck scale. 
\begin{figure}
	\begin{center}
		\includegraphics[scale=0.5]{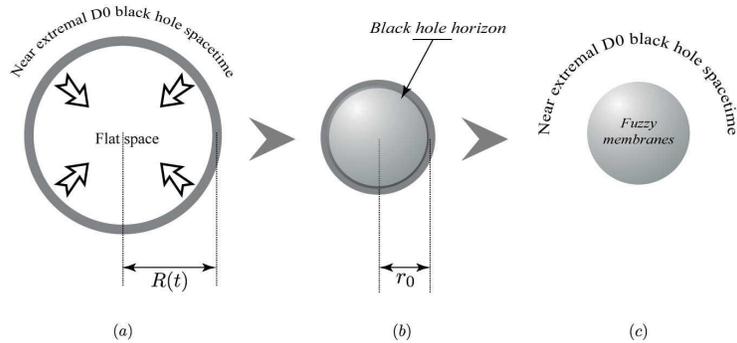} 
	\end{center}
	\caption{The collapse of a thin shell of D0 brane matter as a function of time from (a) to (c); In (b) the radius of the shell $R(t)$ equals that of its horizon at $r_0$ and a D0 black hole emerges.} \label{fig:shellcollapse} 
\end{figure}

Our starting point is a spherical shell of strongly interacting D0 branes in $D$ spacetime dimensions, initially held at rest. We denote the ADM mass of the shell by $M$, $N$ is the number of D0 branes, with $R_0$ being the initial radius. We take $R_0$ large enough so that the horizon size $r_0$ associated with the mass distribution is much less than the initial radius $R_0$. 

We then let go of the shell and have it evolve as a function of proper time $\tau$ until its radius $R(\tau)$ becomes small enough to reveal the horizon at $r_0$ (see Figure~\ref{fig:shellcollapse}). In the context of M(atrix) theory~\cite{Banks:1996vh}, this scenario captures the physics of gravitational collapse and emergence of a Schwarzschild black hole in light-cone M theory with $N$ units of longitudinal momentum.

Initially, we can describe the evolution of the D0 branes within low energy IIA supergravity. Outside the shell, the metric is given by that of the finite temperature D0 black hole 
\begin{equation}
	\label{outsidemetric} ds_E^2=-H^{-7/8} h dt^2+H^{1/8} \left( h^{-1} dr^2+\sum_{i=1}^d dy_i^2+r^2 d\Omega_{8-d}^2\right) 
\end{equation}
with 
\begin{equation}
	\label{outsidedefs} H=1+\frac{k}{L^d r^{7-d}} \ \ \ ,\ \ \ h=1-\frac{r_0^{7-d}}{r^{7-d}}\ ; 
\end{equation}
$k$ and $r_0$ relating to the BPS mass and energy above extremality respectively, and
$L$ denotes the radius of $d$ small circles we may choose to compactify the background on to describe the emergence of a $D=10-d$ dimensional black hole. Inside the shell, the metric is flat. Initially, for $R(\tau)\gg r_0$, the thin shell approximation can be used and the Israel junction condition across the shell tells the story of evolution 
\begin{equation}
	\label{israelcond} \sqrt{1+{\dot{R}^2}}-F^2 H^{7/16} \sqrt{h+\frac{{\dot{R}}^2}{F^2}}=\frac{\pi G_{N} E(R)}{\Omega_{8-d} L^d H^{d/16} R^{7-d}} 
\end{equation}
$E$ is the energy of the shell in the shell's frame,  $F=(7-d+(9+d) H)/(16 H)$, $G_N$ is the ten dimensional gravitational constant, and $\Omega_{8-d}$ is the area of the unit $8-d$ dimensional sphere. For simplicity, we choose to focus onto a window of entropies from $S_{min}\simeq N$ to $S_{max}\simeq N^2$, remembering that the collapse  is an adiabatic process.

Deferring the details of the analysis to the main text, we now summarize the results and present the narrative of black hole formation. As the radius of the shell $R(\tau)$ reaches $r_0$ - the D0 black hole's horizon - we find that the D0 phase in the shell undergoes a phase transition. The partons of the shell rearrange themselves into bound states consisting of spherical D2 branes evenly distributed over the $S^{D-d-2}$. We propose that these fuzzy membranes maintain a size of order the black hole horizon radius $r_0$ through the Myers effect~\cite{Myers:1999ps}. The local mean field D2 flux is of the right order of magnitude to back react on the membranes, while the energy and entropy of the new configuration correctly accounts for the energy and entropy of the emerging D0 black hole! 

This constitutes an explicit realization of Mathur's fuzzball proposal through the process of black hole formation. From the perspective of M(atrix) theory, it describes the collapse of matter into a Schwarzschild black hole in light-cone M theory. The D0 black hole is seen as a boosted Schwarzschild black hole. And since all initial matter configurations in light-cone M theory set up for a collapse are necessarily built out of D0 branes, the conclusions extend to a more general class of problems involving black hole formation and observers falling past Schwarzschild horizons. We track the process through the rest frame of the collapsing shell. Hence, this implies that the horizon of a large finite temperature black hole is not a cross-over line in spacetime that an in-falling observer can harmlessly cross; instead one should think of the Schwarzschild metric ending at around the horizon, replaced inside by a collection of fuzzy membranes. 

The heart of the phenomenon resides in the strong coupling dynamics of $2+1$ dimensional non-commutative Super Yang-Mills (NCSYM), the theory describing the dynamics of the fuzzy membranes~\cite{Seiberg:1999vs}. And we get access to this strongly coupled regime through the holographic duality~\cite{Maldacena:1997re,Gubser:1998bc,Itzhaki:1998dd}. We show that the thermodynamics of fuzzy membranes of spherical topology involves a Gregory-LaFlamme type phase transition~\cite{Gregory:1993vy} that correctly maps onto the point of black hole formation of the D0 system. This transition has a universal character, insensitive to the details of the setup. The implication is that strong coupling dynamics of collapsing matter ``knows'' about the soon to emerge black hole horizon. Hence, looking at the Einstein equation in general, $G_{ab}=8\pi G_N T_{ab}$, we observe: while the left hand side breaks down at Planck scale length scales, we suggest that the energy-momentum tensor on the right hand side - traditionally accorded to a local field theory - changes character at strong coupling at the horizon of the black hole, even when this horizon is at length scales much larger that the Planck scale. It is highly non-trivial that this point of transition on the matter side of the equation knows about the geometrical dynamics on the left side of the equation.

We present the details of our analysis in several stages. In Section 2, we define the $2+1$ dimensional NCSYM theory that lives on fuzzy membranes, or D2 branes with magnetic flux. In Section 3, we describe the strong coupling regime of this theory through the holographic dual picture~\cite{Maldacena:1999mh,Gubser:2004dr,Alishahiha:1999ci} involving the near horizon geometry of the D0-D2 system~\cite{Breckenridge:1996tt}. This allows us to map out the thermodynamics phase diagram of the $2+1$ dimensional NCSYM in Section 4. We present a summary of the relevant phase structure in that section, along with the arguments on how to extend the scaling analysis to the case of fuzzy membranes of spherical topology; but the details of the derivation of the phase diagram are pushed to Appendix A to avoid clutter. We also present an argument through which the fuzzy membranes can maintain a large size in the new black hole phase - tying the discussion to Matrix black holes~\cite{Horowitz:1997fr,Banks:1997hz,Banks:1997cm,Banks:1997tn}. Finally, in the Discussion Section, we present some speculations - mostly with regards to the dynamics of the black hole phase - along with suggestions on how to extend the results to other cases and beyond the thermodynamic scaling analysis we use. 

\section{NCSYM in 2+1 dimensions}

In this section, we define the $U(N_2)$ Non-Commutative Super Yang-Mills theory associated with the dynamics of $N_2$ D2 branes in IIA theory with a magnetic flux on their worldvolume. The Yang-Mills coupling $g_{YM}$ is given by~\cite{Seiberg:1999vs} 
\begin{equation}
	\label{NCSYMgym} g_{YM}^2=\frac{G_s}{l_s} 
\end{equation}
with 
\begin{equation}
	\label{NCSYM} G_s= \frac{g_s B_{x_1 x_2}}{g_{x_1 x_1}} \ \ \ ,\ \ \ G_{x_1 x_1}=G_{x_2 x_2}=\frac{B_{x_1 x_2}^2}{g_{x_1 x_1}} \ \ \ ,\ \ \ \theta^{x_1 x_2}=\frac{2\pi \alpha'}{B_{x_1 x_2}}\ .
\end{equation}
The non-commutative open string coupling of the parent theory is denoted by $\sqrt{G_s}$; and the dynamics lives on a 2+1 dimensional space with diagonal metric $G_{ab}$ and non-commutativity scale $\theta^{x_1 x_2}$ in the two spatial directions $x_1$ and $x_2$. $B_{x_1 x_2}$, $g_{ab}$, $g_s$, and $\alpha'=l_s^2$ are respectively the closed string NSNS B-field, the closed string metric, the closed string coupling, and the string scale\footnote{Note that, as compared to~\cite{Seiberg:1999vs}, in our conventions we have $B^{SW}_{12}=B_{x_ 1 x_2} V_2/(2\pi \alpha')$ where $B^{SW}_{12}$ is the B-field defined in~\cite{Seiberg:1999vs}.}. This theory is to be put on a 2-sphere with the coordinates $x_1$ and $x_2$ of the NCSYM regarded as compact; the area of the 2-sphere is denoted by $V_2$.

At finite temperature $T$, the dimensionless effective coupling is 
\begin{equation}
	\label{geffNCSYM0} g_{eff}^2\equiv \frac{g_Y^2}{T}\ . 
\end{equation}
At strong coupling $g_{eff}^2\gg 1$, one needs to switch to the dual degrees of freedom of IIA supergravity in the background geometry of the D0-D2 system, which we focus on next.

\section{Gravitational dual of NCSYM}

In this section, we look at the holographic dual of $2+1$ dimensional NCSYM theory. This consists of the background geometry cast about the D0-D2 bound system. We will first consider the setting with the D2 branes wrapping a torus. In the next Section, we will show how we can map some conclusions from this setup onto strongly coupled NCSYM on a 2-sphere. We treat here the case relevant to analyzing the emergence of the $D=10$ D0 black hole; and we leave the details of the generalization to $4\leq D<10$ to Appendix A.

\subsection{The D0-D2 background geometry}\label{rawstuff}

The geometry about a D0-D2 system, dual to the strongly coupled regime of the NCSYM of interest, is described by the metric~\cite{Gubser:2004dr} 
\begin{equation}
	\label{metric} ds_{str}^2= H^{-1/2} \left( -h dt^2+D \left( dx_1^2+dx_2^2\right)\right)+H^{1/2} \left( h^{-1}dr^2+r^2 d\Omega_6^2\right)\ , 
\end{equation}
with 
\begin{equation}
	\label{defs} H=1+\frac{q^5}{r^5}\ \ \ ,\ \ \ h=1-\frac{r_0^5}{r^5}\ \ \ ,\ \ \ D=\frac{Q_0^2+Q_2^2}{H^{-1} Q_0^2+Q_2^2}\ . 
\end{equation}
The various constants appearing in these expressions are given by 
\begin{equation}
	\label{Qs} Q_0=N_0 T_0=\frac{N_0}{g_s l_s}\ \ \ ,\ \ \ Q_2=N_2 T_2 V_2=\frac{N_2 V_2}{g_s \left( 2 \pi\right)^2 l_s^3}\ ; 
\end{equation}
with $q^5$ defined through 
\begin{equation}
	\label{q5} \sqrt{Q_0^2+Q_2^2}= \frac{5 V_2 \Omega_6}{2\kappa^2} r_0^{5/2} q^{5/2} \sqrt{1+\frac{q^5}{r_0^5}}\ ; 
\end{equation}
and 
\begin{equation}
	\label{constants} 2\kappa^2=\left( 2\pi\right)^7 g_s^2 {\alpha'}^4 \ \ \ ,\ \ \ \Omega_6=\frac{16}{15}\pi^3\ . 
\end{equation}
This is the spacetime about a bound state of $N_2$ D2 branes and $N_0$ D0 branes. The two dimensional theory is wrapped on a torus, with the directions parallel to the worldvolume of the D2-branes $x_1$ and $x_2$ compactified as in 
\begin{equation}
	\label{compactify} x_1\simeq x_1+\sqrt{V_2} \ \ \ ,\ \ \ x_2 \simeq x_2+\sqrt{V_2}\ . 
\end{equation}
$V_2$ is to eventually be identified with the area of a 2-sphere. The dilaton is given by 
\begin{equation}
	\label{dilaton} e^{2\phi}=g_s^2 H^{1/2} D\ . 
\end{equation}
One has a non-zero NSNS B field 
\begin{equation}
	\label{B} B_{x_1 x_2} = \frac{Q_0}{Q_2} \frac{D}{H}\ ; 
\end{equation}
and the RR fields from the D0 brane and D2 brane charges are\footnote{Note that all the fields have the proper asymptotics for the holographic frame.} 
\begin{equation}
	\label{A1} A_t=-\frac{r_0^{5/2}}{q^{5/2}} \left( 1+\frac{q^5}{r_0^5}\right)^{1/2} \frac{Q_0}{Q H} 
\end{equation}
\begin{equation}
	\label{A3} A_{t x_1 x_2}= - \frac{r_0^{5/2}}{q^{5/2}} \left( 1+\frac{q^5}{r_0^5}\right)^{1/2} \frac{D}{H} \frac{Q}{Q_2} 
\end{equation}
The ADM mass is 
\begin{equation}
	\label{mass} M=\frac{V_2 \Omega_6}{2\kappa^2} \left( 6 r_0^5+5 q^5\right)\ ; 
\end{equation}
while the entropy is given by 
\begin{equation}
	\label{entropy} S =\frac{4\pi V_2 \Omega_6}{2\kappa^2} r_0^6 \sqrt{1+\frac{q^5}{r_0^5}}\ . 
\end{equation}
The $r_0\rightarrow 0$ limit corresponds to the extremal limit. To use this geometry to describe the strong coupling thermodynamics of the NCSYM, we first need to identify the decoupling limit. 

\subsection{Decoupling limit of the D0-D2 system}

The parameters of the NCSYM theory dual to the geometry described by~(\ref{metric}) were listed in~(\ref{NCSYMgym}) and~(\ref{NCSYM}), with $B_{x_1 x_2}$ and $g_{x1 x1}$ given by the asymptotic values of the bulk geometry at $r\rightarrow \infty$ 
\begin{equation}
	\label{asymptotic} g^\infty_{x1 x1}\rightarrow 1 \ \ \ ,\ \ \ B_{x_1 x_2}^\infty=-(2\pi)^2\alpha' \frac{N_0}{N_2 V_2}\ . 
\end{equation}
We then have 
\begin{equation}
	\label{NCSYMparam} \label{paramsNCSYM} G_{x_1 x_1}=G_{x_2 x_2}=(2\pi)^4 \frac{N_0^2}{N_2^2} \frac{{\alpha'}^2}{V_2^2} \ \ \ ,\ \ \ \theta^{x_1 x_2}=\frac{1}{2\pi}\frac{N_2}{N_0} V_2 \ \ \ ,\ \ \ {G_s}=(2\pi)^2 \frac{g_s}{V_2} \frac{N_0}{N_2} \alpha'\ . 
\end{equation}
This implies that the D2 brane worldvolume coordinates do not commute as in $[{\hat{x}}^1, {\hat{x}}^2]=(i/2\pi) (N_2/N_0)$ where we write the algebra in terms of the rescaled coordinates ${\hat{x}}^a=x_a/V_2$. Hence, the effective dimensionless NCSYM coupling at temperature $T$ is given by 
\begin{equation}
	\label{geffNCSYM} g_{eff}^2\equiv \frac{g_Y^2}{T}= \frac{G_s}{l_s T}=\frac{N_0}{N_2}(2\pi)^2 \frac{g_s l_s}{V_2 T}\ . 
\end{equation}

To zero onto the energy scales of NCSYM, we take $\alpha'\rightarrow 0$ with 
\begin{equation}
	\label{gslimit} g_s\sim l_s^3\ , 
\end{equation}
This keeps the Yang-Mills coupling $g_{YM}^2\sim g_s/l_s^3$ fixed. We also need 
\begin{equation}
	\label{decoupling} e^\phi\sim \mbox{finite} \ \ \ ,\ \ \ ds_{str}^2\sim\alpha' \ \ \ ,\ \ \ B_{12}\sim 1/\alpha' 
\end{equation}
The first statement assures that one has a perturbative string theory in the resulting background geometry; the second assures decoupling from gravity; and the last assures non-commutation of $x_1$ and $x_2$ in the boundary NCSYM theory. We then arrive at the following scaling relations 
\begin{equation}
	\label{decoupling2} H^{1/2} D g_s^2\sim 1 \ \ \ ,\ \ \ H^{1/2} r^2\sim \alpha' \ \ \ ,\ \ \ H^{-1/2} D V_2 \sim \alpha' 
\end{equation}
These set of conditions have one solution of interest given by\footnote{The conditions lead to two possible solutions, one of which corresponds to low enough energies that the field theory is commutative 2+1 dimensional Yang-Mills. This energy regime is too restrictive for the problem at hand; we later find that the energy window of relevance to the shell collapse problem includes the scale of non-commutativity.} 
\begin{equation}
	\label{decouplingres} g_s\sim l_s^3 \ \ \ ,\ \ \ r\sim \alpha' \ \ \ ,\ \ \ V_2\sim {\alpha'}^2\ . 
\end{equation}

Hence, we define the more convenient finite parameters 
\begin{equation}
	\label{definitions} g\equiv \frac{g_s}{l_s^3} \ \ \ ,\ \ \ v_2\equiv \frac{V_2}{{\alpha'}^2} \ \ \ ,\ \ \ u\equiv \frac{r}{\alpha'} 
\end{equation}
which we use to rewrite the needed relations from Section~\ref{rawstuff} in the decoupling regime of interest 
\begin{equation}
	\label{q5limit} q^5=\frac{3}{2} (2\pi)^4 \frac{N_0 g}{v_2} {\alpha'}^3 
\end{equation}
with 
\begin{equation}
	\label{Heq} H=\frac{6 Q_0}{5 E}=\frac{6 N_0}{5 g E} \frac{1}{{\alpha'}^2} 
\end{equation}
The energy above extremality $E\equiv M-M_{BPS}$ is 
\begin{equation}
	\label{energyphase} E=\frac{4}{5 (2\pi)^4} \frac{v_2}{g^2} u_0^5 
\end{equation}
with $u_0\equiv r_0/\alpha'$. Note also that in the decoupling limit $M_{BPS}$ of the D0-D2 system is dominated by the BPS mass of the D0 branes 
\begin{equation}
	\label{MBPS} M_{BPS}=\sqrt{M_{D0}^2+M_{D2}^2}\rightarrow M_{D0}=\frac{1}{{\alpha'}^2}\frac{N_0}{g}\rightarrow \infty\ . 
\end{equation}
And the entropy is given by 
\begin{equation}
	\label{entropyphase} S^2=\frac{8}{75 (2\pi)^2} \frac{v_2 N_0}{g^3} u_0^7 
\end{equation}
Finally, we need the important factor 
\begin{equation}
	\label{DoverH} \frac{D}{H}=\frac{1}{1+\varepsilon}\equiv \Delta 
\end{equation}
where we have defined 
\begin{equation}
	\label{epsilon} \varepsilon\equiv \frac{6 N_2^2 v_2^2}{5 (2\pi)^4 N_0 g E} 
\end{equation}
The regime of interest to the collapsing shell problem corresponds to $\varepsilon\ll 1$ as we will discover below. Part of the magic resides in the fact that in this regime the thermodynamics is describing D0 branes smeared in two directions; a configuration that - through a Gregory-LaFlamme transition - perfectly matches onto the initial phase of a shell of collapsing D0 branes.

For completeness, we also write the map between energy scale in the boundary NCSYM theory and the bulk holographic coordinate $u$. To do this, one focuses on null geodesics in the $u-t$ plane, writing $t\rightarrow 1/E$ where $E$ is energy scale in the NCSYM theory; one then gets 
\begin{equation}
	\label{uvir} T\simeq \sqrt{\frac{v_2}{N_0 g}} u^{3/2}\ . 
\end{equation}
Hence, as expected, high temperatures correspond to the bulk region near the $u\rightarrow \infty$.

\section{The black hole formation narrative}

The task is to explore the strong coupling regime of the theory defined in Section 2. This is achieved by looking at the near horizon geometry of a D0-D2 system, and mapping the physics from this geometrical viewpoint onto the NCSYM theory through the holographic duality. One immediate problem however is that the theory we are interested in lives on a 2-sphere; this is so as to tie onto the shell collapse problem later. And we want the analysis in the rest frame of the D0-D2 system since the conclusions are to be applied to the rest frame of the in-falling shell. But spherical D2 branes with D0 charge are not a priori stable configurations and the corresponding supergravity geometry would generically be time dependent ({\em i.e.} see equation~(\ref{israelcond})). The alternative is to consider the D0-D2 system with the D2 branes wrapping an artificially stabilized torus of area equal to the area $V_2$ of the corresponding sphere. However, the toroidal configuration would conserve wrapped D2 brane charge and avoids phase transitions that otherwise can exist for spherical D2 branes. For a spherical configuration, the D2 brane charge is multipole and hence is not conserved: a spherical D2 brane can disappear whereas one wrapped on a torus cannot. 
\begin{figure}
	[t] 
	\begin{center}
		\includegraphics[scale=0.7]{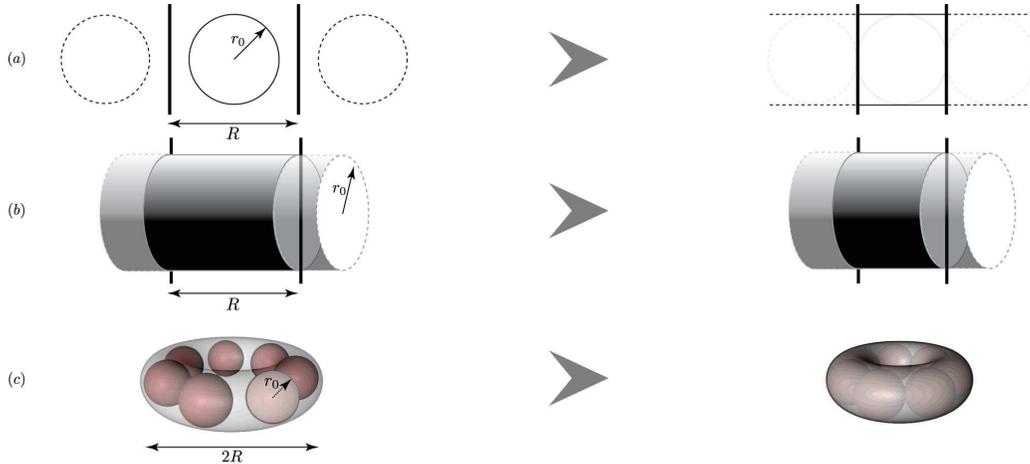} 
	\end{center}
	\caption{Cartoons of various geometrical phase transitions: (a) A black hole of horizon size $r_0$ is on a circle of size $R$; the dotted circles denote the images arising from the periodic boundary conditions; as the box size decreases to $R\simeq r_0$, a Gregory-LaFlamme transition to a phase more uniform along the circle is expected. (b) A black string is wrapped on the circle; as the circle shrinks in size to $r_0\gg l_s$, no phase transition is expected; the wrapped string charge is conserved. (c) A ring of black holes shrinks in size; when $R\simeq r_0$, the horizons touch (much like in (a)) and a phase transition is again expected; in this case brane charge of multipole order need not be conserved.} \label{fig:GL} 
\end{figure}

The thermodynamic phase transition of interest is the well-known Gregory-LaFlamme transition~\cite{Gregory:1993vy}. Figure~\ref{fig:GL}(a) is a cartoon of the physics involved: when a black hole's horizon becomes big enough to probe the size of a compact direction, the hole would want to decay into a configuration `smeared' along the compact direction. Effectively, the horizon of the black hole touches the horizon of its images in the compact direction and the images can be thought of as fusing together: the bottom line is that one cannot fit a large black hole in too small of a box. In the case of a brane wrapping a circle as shown in Figure~\ref{fig:GL}(b), a similar argument is not present and the wrapped brane charge is properly conserved. Figure~\ref{fig:GL}(c) illustrates the scenario of interest to us. A ring of black objects gravitationally collapses, shrinking the ring's radius; as the radius reaches the size of the horizon, the horizon surfaces start fusing and would be unstable toward forming the more uniform configuration of a smeared ring, much like the Gregory-LaFlamme argument of Figure~\ref{fig:GL}(a). This toy example carries over to higher dimensions, particularly to the case where D0 branes get smeared on a $S^{D-d-2}$ through the process of collapse. We will see that this can be viewed as the D0 branes forming bound states consisting of D0-D2 systems. Of the sequence available in IIA theory, it is known that the D0-D4 system is only marginally bound, and the D0-D6 through the D0-D8 cost more energy than the sum of the masses of their constituents~\cite{Taylor:1997ay}-\cite{Friess:2005tz}. However, the D0-D2 is a truly bound state: the D0's lower their energy by creating a network with D2 brane charge. We will find that indeed the D0-D2 matches onto the evolution of the collapsing shell through a Gregory-LaFlamme transition of the type described in Figure~\ref{fig:GL}(c), correctly accounting for the entropy and energy of the new D0 black hole - and according an intriguing universal character to the black hole formation process. For example, carrying out a similar analysis for the D0-D4 system instead, we have found that the the D0-D4 phase cannot connect to the initial D0 phase. 

Hence, inspired by the discussion above, we capture the interesting transition points in the spherical case by using the static toroidal background geometry of the D0-D2 system, and looking for the point where the horizon size competes with the size of the torus. While we should not expect that we would be able to identify {\em precisely} the phase transitions, we do expect that the scaling relations between the thermodynamic parameters at the transition points are captured correctly. We will next adopt this hybrid strategy to map out the phase diagram of NCSYM living on a 2-sphere. This can then be used to describe the phases of the collapsing shell matter in the rest frame of the shell.

\subsection{The phase diagram}

Starting from the geometry described by~(\ref{metric}), and considering the discussion above about geometrical transitions of the Gregory-LaFlamme type, we then map out the phase diagram of the shell matter. In this process, we encounter various regimes requiring descriptions in dual degrees of freedom. Applying T-dualities and lifts to M theory as needed, we arrive at a consistent picture for the phase structure that we now summarize. The details of the computations are in Appendix A. 
\begin{figure}
	[t] 
	\begin{center}
		\includegraphics[scale=0.5]{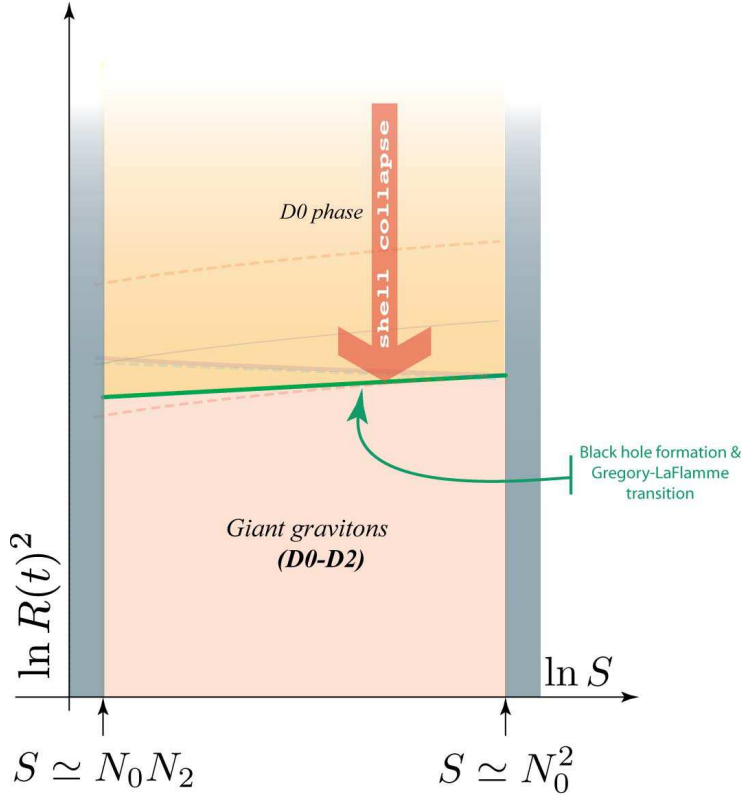} 
	\end{center}
	\caption{The phase diagram describing the matter making up the collapsing shell in the rest frame of shell. The faint lines on the diagram correspond to various duality transformations explained in Appendix A.}\label{fig:simplephase} 
\end{figure}

Figure~\ref{fig:simplephase} shows the resulting phase diagram of NCSYM on a sphere in the entropy window $N_0 N_2<S<N_0^2$ (note that the number $N$ of D0 branes in the initial phase gets mapped to $N_0$)\footnote{For entropies $S\sim N_0^2$, the effective coupling in the $0+1$ dimensional NCSYM is small; the interactions between the D0 branes are weak with the equation of state fixed through dimensional analysis $E\sim N_0^2 T\sim T\,S\Rightarrow S\sim N_0^2$; in this state, one may not be able to arrange an initial cohesive shell of D0 branes without having them fly apart. For entropies $S<N_0 N_2$, the initial phase of D0's is again of different character, one with a natural description directly in M theory - as can be seen in the detailed computations in Appendix A.}. Given that the collapse process is adiabatic, we choose to present the results in terms of entropy instead of free energy. The horizontal axis is log of the entropy, while the vertical axis is log of the radius of the shell squared or equivalently $v_2$. The upper part of the diagram describes a phase of D0 branes of which the collapsing shell is initially made of, before the black hole horizon emerges. The equation of state of this phase is given by 
\begin{equation}
	\label{d0eos0} E_{D0}\simeq N^{-7/9} g^{1/3} S^{14/9}\ . 
\end{equation}
The ADM energy as measured by an observer at infinity is $M_{ADM}=Q_0+E_{D0}$, with $Q_0=M_{BPS}^{(D0)}$ given by~(\ref{Qs}). The phase at the bottom of the diagram is described by the equation of state of the D0-D2 system 
\begin{equation}
	\label{eosd0d2} E_{D0D2}\simeq \frac{v_2^{2/7} g^{1/7}}{N_0^{5/7}} S^{10/7}\ . 
\end{equation}
By the holographic duality, the energy of the shell as measured in the rest frame of the shell is $M_{BPS}+E_{D0D2}$ with $M_{BPS}$ given by~(\ref{MBPS}), which is the same as $Q_0$ in the decoupling limit. This phase describes fuzzy membranes.

The collapse process then traces a vertical line from top to bottom as depicted in the Figure. A Gregory-LaFlamme transition of the type described in Figure~\ref{fig:GL}(c) is indicated by a solid line. It is at 
\begin{equation}
	\label{maintransition} v_2\simeq g^{4/7} E_{D0D2}^{2/7}\ . 
\end{equation}
At this point, the D0's get smeared over the $S^{8}$ and bind together to form D2 branes with magnetic flux. We will argue in the next subsection that the collapse of the shell can be stopped by the Myers dielectric effect that fuzzy membranes are subject to. Note that this transition point is independent of $N_0$ and $N_2$. For $D<10$ dimensions, {\em i.e.} with the initial phase of D0's smeared over $d$ circles of size $\mathcal{L}=L/\alpha'$, we find the Gregory-LaFlamme transition point at (see Appendix A for the details) 
\begin{equation}
	\label{maintransitioninD} E_{D0D2}\simeq \frac{g^2 v_2^{\frac{7-d}{2}}}{\mathcal{L}^d}\ . 
\end{equation}
Focusing back on the $d=0$ case for simplicity, we can convert~(\ref{maintransition}) using~(\ref{eosd0d2}) to a statement with respect to the entropy 
\begin{equation}
	\label{maintransitionS} v_2\simeq N_0^{-2/9} g^{2/3} S^{4/9}
\end{equation}
which, writing $v_2\rightarrow r_0^2$, we recognize as the entropy-horizon radius relation for the D0 black hole!
The entropy $S$ is the same for the observer at infinity and the one sitting at the horizon. Hence, we can use~(\ref{d0eos0}) to write a relation for the energy of the new phase as seen by the observer at infinity; this gives 
\begin{equation}
	\label{maintransition3} v_2\simeq g^{4/7} E_{D0}^{2/7} 
\end{equation}
using $N=N_0$, and this is naturally the right relation for the D0 black hole. The asymptotic charges also match, having $N=N_0$ D0 charge and zero net D2 brane charge. For the case involving a $D=10-d$ dimensional D0 black hole, noting that the gravitational constant is $G_N\simeq g_s^2 {\alpha'}^4/L^d$, we recognize the relation~(\ref{maintransitioninD}) with $E_{D0D2}=E_{D0}$ as the transition point where a $D$ dimensional D0 black hole horizon emerges. All this assuming that the fuzzy membranes of the new phase maintain their large radii $\sqrt{V_2}$ at the size of the black hole horizon. We will justify this assumption in the next subsection. Furthermore, applying the same analysis and arguments with respect to, say, a D0-D4 system - {\em i.e.} considering the smearing of the D0's on the $S^8$ leading to spherical fuzzy D4's distributed over the $S^8$ - can be shown to lead to inconsistencies. Hence, the picture involving D0's coalescing into fuzzy membranes in a consistent thermodynamic framework is indeed highly non-trivial.

In summary, as the D0 shell collapses, at the point it is about to reveal the horizon of the D0 black hole, the matter in the shell undergoes a phase transition: the D0 branes form bound systems of D2 branes. According to the spherical symmetry of the initial conditions, these fuzzy 2-branes must cover the $S^{D-d-2}$ that is the horizon of the emerging hole. And their typical size at that point is the size of the horizon. The geometrical description inside the hole then breaks down and it is replaced by a picture involving a soup of D0 branes bound into fuzzy spheres. We will next present an argument that suggests that these fuzzy membranes remain large through the Myers effect. 

\subsection{The phase inside the hole}

At the black hole formation transition point, the degrees of freedom of the matter in the shell have been converted from D0 branes to fuzzy membranes, or equivalently bound states of D0 branes forming D2's. Initially, the size of the emerging fuzzy membranes is the size of the horizon, and from symmetry we would expect a uniform distribution on the $D-d-2$ dimensional sphere that is the emerging horizon. While the dynamics is rather complicated, it is however possible to get a good qualitative feel of this new phase when one realizes that all the ingredients needed for the Myers dielectric effect are present in this soup.

In general, the dielectric effect in question involves a 4-form flux of strength $f$ in flat space polarizing $N_0$ D0 branes into a D2 brane of size~\cite{Myers:1999ps} 
\begin{equation}
	\label{Myerssize} b\simeq \alpha' f N_0\ . 
\end{equation}
The resulting fuzzy membrane is associated with a scale of non-commutativity given by 
\begin{equation}
	\label{noncommscale} [x^1,x^2]\sim i \alpha' f x^3\ . 
\end{equation}
In our case, we expect a number of bound D2 systems to form, stretching in the various angular directions of the $S^{D-d-2}$. This implies that, while the net D2 brane charge of the system is zero, there will be a mean field multipole 4-form field strength in the region inside the horizon for certain orderly arrangements of the D2 branes. This is a common phenomenon in many condensed matter systems involving polarization of molecules. We can estimate the size of this mean field flux due to back-reaction. 

In our conventions, $f$ has units of inverse length. There are two length scale in the problem: the string scale $l_s$ and the average size of the fuzzy membranes $V_2$. We expect that the leading contribution to $f$ is dipole and would be larger for larger fuzzy membranes. This implies that $f\propto \sqrt{V_2}/\alpha'\propto \sqrt{v_2}$ as the leading term which would be finite in the decoupling limit. We may however have a dimensionless coefficient to this expression involving\footnote{$g_s$ cannot appear without making the resulting flux zero or infinite in the decoupling limit.} the numbers $N_0$ and $N_2$. We can then write 
\begin{equation}
	\label{estimate} f_{mean}\sim q(N_0,N_2) \sqrt{v_2} 
\end{equation}
for some unknown dimensionless function $q(N_0,N_2)$. Using~(\ref{Myerssize}), a field strength of this size sustains fuzzy membranes of radius 
\begin{equation}
	\label{estimate2} b\simeq \alpha' N_0\ q(N_0,N_2) \sqrt{v_2}\ , 
\end{equation}
associated with a scale of non-commutativity given by~(\ref{noncommscale}), or 
\begin{equation}
	\label{noncommscale2} \alpha' f b\sim {\alpha'}^2 N_0 q^2 v_2\ , 
\end{equation}
where we set $x^3\sim b$ given that $x_1^2+x_2^2+x_3^2\sim b^2$, {\em i.e.} we are looking near one of the poles of the spherical membrane~\cite{Myers:1999ps}. But comparing this to $\theta$ in~(\ref{NCSYMparam}), we find 
\begin{equation}
	\label{balancenc} \alpha' f b \sim \frac{N_2}{N_0} v_2 {\alpha'}^2\Rightarrow q\simeq \frac{\sqrt{N_2}}{N_0} 
\end{equation}
This implies 
\begin{equation}
	\label{sizeeq} b\simeq \sqrt{N_2 V_2}\ \ \ \ \ \ \mbox{and}\ \ \ \ \ \ f_{mean}=\frac{\sqrt{N_2 v_2}}{N_0}\ . 
\end{equation}
For $b\sim \sqrt{V_2}$, this requires that
$N_2$ is of order one, as assumed throughout the analysis. To understand why forming a few D2 branes would be favored over forming a large stack of $N_2\gg 1$ D2's, one needs to look into the details of the dynamics of the new phase. We defer an  in-depth analysis to a future work and propose that these observations suggest that the Myers dielectric effect can account for large fuzzy membranes of the size of the horizon with $N_0\gg N_2$; the mean field flux $f_{mean}$ that is needed being parametrically small with large $N_0$.

This picture fits very well with the rest of the narrative, and the degrees of freedom of the black hole can be viewed as being fuzzy membranes. This also ties in well with the story of Matrix black holes~\cite{Horowitz:1997fr,Banks:1997hz,Banks:1997cm,Banks:1997tn}. In that setting, it was proposed to model the finite temperature black hole as a gas of D0 branes in a ball formation of the size of the horizon. One problem in that picture was that one had to assume that these D0 branes behave like distinguishable partons, so as to account for the black hole entropy correctly. In~\cite{Banks:1997tn}, it was proposed that this may arise because the D0's may be threaded to D2 branes. We see here that we essentially have realized this proposal through the process of gravitational collapse.

\section{Discussion} 

In this work, we show that a phase of strongly interacting D0 branes - arranged in a spherical formation that is gravitationally collapsing - undergoes a phase transitions as the shell's radius reaches the horizon of the mass distribution. This phase transition is a strong coupling phenomenon, arising in a non-commutative non-local theory. The candidate emerging phase appears to be a soup of fuzzy membranes whose energy and entropy match onto those of the D0 black hole - if these membranes can sustain their large initial size. We argue that the dielectric Myers effect provides such a mechanism through back-reaction. The entire discussion can be cast in the framework of M(atrix) theory to describe the process of forming a boosted Schwarzschild black hole in light-cone M theory.

It is important to emphasize the special role played by fuzzy D2 branes in this process in {\em arbitrary} spacetime dimension $D$: indeed, the same phenomenon is absent when one considers fuzzy D$p$ branes with $p\geq 4$. This connects well with the picture that the D0's are forming bound states at the point of horizon emergence, since it is known that, in the D0-D$p$ sequence, only the $p=2$ case leads to a state with lower energy than the sum of the constituents~\cite{Taylor:1997ay,Sheinblatt:1997nt,Witten:2000mf,Mihailescu:2000dn}\footnote{The D0-D4 system is marginally bound; and candidates for D0-D6 or D0-D8 in the literature have total energy greater than the sum of the BPS masses of the constituent D-branes.}.

The new phase of fuzzy membranes entails interesting new dynamics that needs to be further explored~\cite{Kabat:1997im,Chepelev:1998sm,Lee:2004kv}. In particular, one may expect an ordered, lattice-like distribution of the D2 branes so as to have a net non-zero dipole charge inside the configuration. Indeed, we are able to account for the entropy of the entire black hole through the finite temperature excitations of the individual D2 branes - with apparently no relevant contribution from the center of mass dynamics of the D2 branes; this suggests that the phase space for the center of mass dynamics is restricted, perhaps because of the orderly arrangement of the branes. Our analysis gives a plausibility argument for having the membranes maintain large size through the Myers effect; one would like to expand on this line of thought in greater detail - in the process identifying the reason why the number of D2's generated is of order one. Understanding these points can also lead to other interesting checks and extensions, such as the unravelling of the Hawking evaporation phenomenon. Indeed, our initial attempts at probing the dynamics of the soup of fuzzy membranes using the DBI action~\cite{Future} suggest rich dynamics possibly leading to a process of quantum tunneling for black hole evaporation - along the line recently discussed in the literature~\cite{Parikh:1999mf}\footnote{Interestingly, in the work of~\cite{Alberghi:2001cm}, part of the picture of membranes generated through gravitational collapse has been suggested in a non-stringy setting.}. We hope to report on these issues in a future work.

It would also be interesting to extend this picture to other black holes, including rotating ones. It is possible that the fuzzy membranes of the D0 black hole we considered get replaced with other exotic partons, such as giant gravitons~\cite{McGreevy:2000cw} when applied to other settings with more types of charges.

\section{Appendices}

\vspace{0.5in} {\bf \large Appendix A: Deriving the phase structure}\vspace{0.5in}

In this appendix, we present the details of the analysis leading up to the phase diagram shown in Figure~\ref{fig:simplephase}. We first consider the case corresponding to the formation of a 10D D0 black hole, {\em i.e.} the holographic dual includes the 10D IIA supergravity D0-D2 background geometry~(\ref{metric}). Later, we generalize to the case of a $4\leq D < 10$ D0 black hole by smearing the geometry on transverse circles. 
\begin{figure}
	[t] 
	\begin{center}
		\includegraphics[scale=0.7]{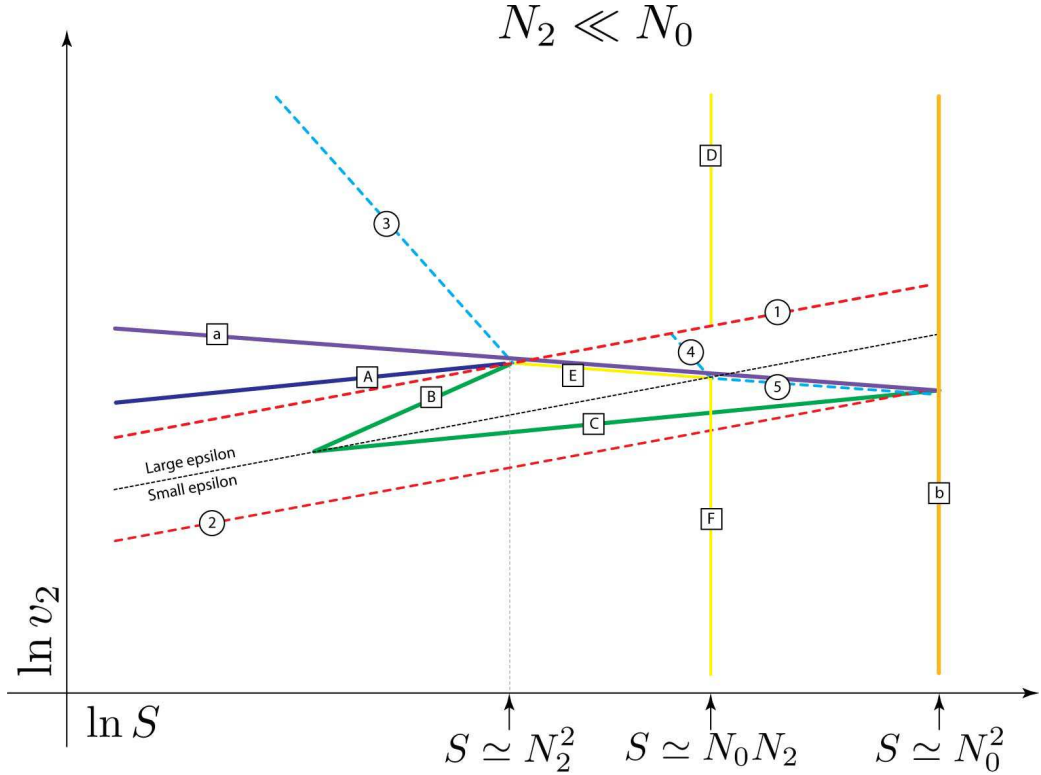} 
	\end{center}
	\caption{The phase diagram of $2+1$ dimensional NCSYM on a 2-sphere of radius $\sqrt{V_2}=\sqrt{v_2}\alpha'$. The labels on the various curves are referenced in the main text. Dotted lines denote duality transformations, whereas solid lines are expected to describe phase transitions. Curve C is the main black hole formation transition of interest in this work.} \label{fig:phasediagram} 
\end{figure}

In our problem, $N_0$, the number of D0 branes in the D0-D2 system, is to eventually be mapped onto the number $N$ of D0 branes making up the collapsing shell. We take $N_0=N$ large and focus on the regime where $N_0\gg N_2$, where $N_2$ is the number of D2 branes. Furthermore, the collapsing shell will be of spherical shape, whereas the thermodynamics analysis will focus on D0 branes bound into D2 branes wrapped on a torus. We expect that the phase transitions due to finite size effects will be qualitatively similar between the spherical and toroidal cases as described in detail in the main text; the differences would be seen in numerical coefficients for the exact points of phase transitions. In the forthcoming, we follow the approach used in~\cite{Martinec:1998ja,Sahakian:1999gj}.

As mentioned earlier, IIA string theory in the background of the near horizon geometry of~(\ref{metric}) is dual to the strongly coupled $2+1$ dimensional NCSYM of interest. However, this geometry is subject to several conditions that delineate its regime of validity. One needs to have the compact torus larger than the string scale 
\begin{equation}
	\label{Tdualdef} \left.g_{x1 x2}\right|_{horizon}\gg \alpha'\ , 
\end{equation}
This condition becomes (curves 2 and 1 in Figure~\ref{fig:phasediagram}) 
\begin{eqnarray}
	\label{Tdual} v_2 \gg g^{1/2} E^{1/2} N_0^{-1/2}\ \ \ ,\ \ \ &\mbox{ for }&\varepsilon \ll 1 \nonumber \\
	v_2 \ll N_0^{3/2} N_2^{-2} g^{1/2} E^{1/2} \ \ \ ,\ \ \ &\mbox{ for }&\varepsilon \gg 1 
\end{eqnarray}
with $\varepsilon$ defined in equation~(\ref{epsilon}). Otherwise, we need to look at the T-dual geometry on the 2-torus (see Appendix C for the details). The Gregory-LaFlamme stability condition on the 2-torus maps onto the statement 
\begin{eqnarray}
	\label{GLdef} \left.g_{x_1 x_1}\right|_{horizon}<\left.g_{\Omega_6} \right|_{horizon} 
\end{eqnarray}
or (curves C and B in Figure~\ref{fig:phasediagram}) 
\begin{eqnarray}
	\label{GL} v_2 < g^{4/7} E^{2/7} \ \ \ ,\ \ \ &\mbox{ for }&\varepsilon \ll 1 \nonumber \\
	v_2 > N_0^{5/3} N_2^{-10/3} g^{1/3} E \ \ \ ,\ \ \ &\mbox{ for }&\varepsilon \gg 1 
\end{eqnarray}
The first of these two is the main transition point of black hole formation in Figure~\ref{fig:simplephase}. In the T-dual geometry, the Gregory-LaFlamme condition is (curve A in Figure~\ref{fig:phasediagram}) 
\begin{equation}
	\label{GLT} \left.\frac{g_{x_1 x_1} {\alpha'}^2}{g_{x_1 x_1} g_{x_2 x_2}+B_{x_1 x_2}^2}< g_{\Omega_6}\right|_{horizon} \Rightarrow u_0^2>v_2 \left( \frac{N_2}{N_0}\right)^2 \Rightarrow v_2< N_0^{10/7} N_2^{-10/7} g^{4/7} E^{2/7} \ \ \ ,\ \ \ \mbox{ for all }\varepsilon\ . 
\end{equation}
One also needs the dilaton to be small, or 
\begin{equation}
	\label{strongcouplingdef} \left.e^\phi\right|_{horizon}\ll 1 
\end{equation}
or (curves 5 and 4 in Figure~\ref{fig:phasediagram}) 
\begin{eqnarray}
	\label{strongcoupling} E> g^{1/3} N_0 \ \ \ ,\ \ \ &\mbox{ for }&\varepsilon \ll 1\nonumber \\
	v_2>N_0^{5/4} N_2^{-1} g^{3/4} E^{-1/4}\ \ \ ,\ \ \ &\mbox{ for }&\varepsilon \gg 1\ ; 
\end{eqnarray}
or else we lift to M-theory. In the T-dual picture, this condition of lift to M theory becomes (curve 3 in Figure~\ref{fig:phasediagram}) 
\begin{equation}
	\label{strongcoupling2} E>N_2^{4/3} N_0^{-1/3} g^{1/3} \ \ \ ,\ \ \ \mbox{ for all }\varepsilon\ . 
\end{equation}
And finally, one needs the curvature scale of the geometry at the horizon to be very small, or (curve a in Figure~\ref{fig:phasediagram}) 
\begin{equation}
	\label{curvature} \left. g_{\Omega_6}\right|_{horizon}\gg \alpha' \Rightarrow \frac{N_0^5 g^3}{v_2^4}\gg E \ \ \ ,\ \ \ \mbox{ for all }\varepsilon 
\end{equation}
for both main and T-dual pictures.

At strong coupling, we lift to M-theory and we the need to consider Gregory-LaFlamme localization transitions on the eleventh direction; these stability conditions become (curves F and E in Figure~\ref{fig:phasediagram}) 
\begin{eqnarray}
	\label{mloc} v_2< N_0^{-5/2} g^{-1/2} E^{7/2}\ \ \ ,\ \ \ &\mbox{ for }&\varepsilon \ll 1 \nonumber \\
	v_2>N_0^{5/4} N_2^{-5/4} g^{3/4} E^{-1/4} \ \ \ ,\ \ \ &\mbox{ for }&\varepsilon \gg 1\ ; 
\end{eqnarray}
and starting from the T-dual geometry, these become (curves F and D in Figure~\ref{fig:phasediagram}) 
\begin{equation}
	\label{mlocT} v_2< N_0^{5/2} N_2^{-5} g^{-1/2} E^{7/2}\ \ \ ,\ \ \ \mbox{ for all }\varepsilon\ . 
\end{equation}

Under T-duality transformation or M-lift, the equation of state of the phase remains unchanged and is given by~(\ref{eosd0d2}). Using this equation of state, we convert all these transition curves to statements in terms of entropy instead of energy: 
\begin{itemize}
	\item The T duality point (curves 2 and 1 ) 
	\begin{eqnarray}
		\label{tdualS} v_2> N_0^{-1} g^{2/3} S^{5/6}\ \ \ ,\ \ \ &\mbox{ for }&\varepsilon \ll 1 \nonumber \\
		v_2< N_0^{23/12} N_2^{-7/3} g^{2/3} S^{5/6} \ \ \ ,\ \ \ &\mbox{ for }&\varepsilon \gg 1\ . 
	\end{eqnarray}
	
	\item The Gregory-LaFlamme condition (curves C and B) 
	\begin{eqnarray}
		\label{GLS} v_2<N_0^{-2/9} g^{2/3} S^{4/9}\ \ \ ,\ \ \ &\mbox{ for }&\varepsilon \ll 1 \nonumber \\
		v_2> N_0^{4/3} N_2^{-14/3} g^{2/3} S^2 \ \ \ ,\ \ \ &\mbox{ for }&\varepsilon \gg 1 
	\end{eqnarray}
	
	\item The Gregory-LaFlamme in the T dual geometry (curve A) 
	\begin{equation}
		\label{GLST} v_2<N_0^{4/3} N_2^{-14/9} g^{2/3} S^{4/9} \ \ \ ,\ \ \ \mbox{ for all }\varepsilon\ . 
	\end{equation}
	
	\item Small dilaton (curves 5 and 4) 
	\begin{eqnarray}
		\label{smallcouplingS} v_2>N_0^6 g^{2/3} S^{-5}\ \ \ ,\ \ \ &\mbox{ for }&\varepsilon \ll 1 \nonumber \\
		v_2> N_0^{4/3} N_2^{-14/15} g^{2/3} S^{-1/3}\ \ \ ,\ \ \ &\mbox{ for }&\varepsilon \gg 1 
	\end{eqnarray}
	
	\item Small dilaton in the T dual geometry (curve 3) 
	\begin{equation}
		\label{smallcouplingST} v_2>N_0^{4/3} N_2^{14/3} g^{2/3} S^{-5}\ \ \ ,\ \ \ \mbox{ for all }\varepsilon\ . 
	\end{equation}
	
	\item Small curvature at the horizon (curve a) 
	\begin{equation}
		\label{curvaturewithS} v_2<N_0^{4/3} g^{2/3} S^{-1/3}\ \ \ ,\ \ \ \mbox{ for all }\varepsilon 
	\end{equation}
	in both main and T-dual geometries.
	
	\item Localization on the eleventh direction (curves F and E) 
	\begin{eqnarray}
		\label{mlocS} S>N_0 \ \ \ ,\ \ \ &\mbox{ for }&\varepsilon \ll 1 \nonumber \\
		v_2>N_0^{4/3} N_2^{-7/6} g^{2/3} S^{-1/3}\ \ \ ,\ \ \ &\mbox{ for }&\varepsilon \gg 1\ . 
	\end{eqnarray}
	
	\item Localization on the eleventh direction in the T dual geometry (curves D and F) 
	\begin{equation}
		\label{mlocST} S>N_2\ . 
	\end{equation}
\end{itemize}

Hence we can now chart the phase diagram for all entropies and shell radii. The dynamics of interest however involves $N_2\sim 1$ and $N_2\ll N_0$ as discussed in the main text. This means that the left side of Figure~\ref{fig:phasediagram} beyond $S\sim N_2^2$ is irrelevant since entropies would be small enough to correspond to a D0 black hole of Planckian size. The region between $S\sim N_2^2$ and $S\sim N_0 N_2$ is still interesting and always involves eleven dimensional dynamics. We defer a detailed analysis of this region of the phase diagram but present the structure of the diagram for completeness and future reference. Focusing on the region between $S\sim N_0 N_2$ and $S\sim N_0^2$, a careful inspection of the overlapping regimes quickly leads to the simplified phase diagram shown in Figure~\ref{fig:simplephase}.

\vspace{0.3in} {\bf Diverse dimensions} \vspace{0.3in}

In this section, we extend the previous analysis to the cases where the shell is collapsing in less than 10 dimensions. We achieve this by smearing the D0-D2 background geometry in the transverse directions by replacing the Laplacian function by 
\begin{equation}
	\label{sumsemar} \sum_n \frac{1}{((x - n L)^2 + (x^i)^2)^{n/2}}\rightarrow \int_{\infty}^{\infty} \frac{dy}{L} \frac{1}{((x-y)^2+(x^i)^2)^{n/2}}= \sqrt{\pi}\frac{\Gamma(\frac{n-1}{2})}{\Gamma(\frac{n}{2})}\frac{1}{L}\frac{1}{((x^i)^2)^{\frac{n-1}{2}}} 
\end{equation}
where $n=5$ in our case. Note that because one is mapping a toroidal analysis onto a spherical phase structure, we can carry out this smearing at most for four transverse dimensions. Beyond that, the system will become sensitive to the topology and one would need to chart the phase structure for the spherical topology directly. This being an unstable configuration, the analysis would become considerably more involved.

Smearing the D0-D2 geometry in both $H$ and $h$, one finds that the horizon location $u_1\equiv r_1/\alpha'$ is now at 
\begin{equation}
	\label{newhor} \frac{u_0^5}{\mathcal{L}^d u_1^{5-d}}\simeq 1 
\end{equation}
with $E\sim v_2 u_0^5/g^2\sim v_2 \mathcal{L}^d u_1^{5-d}/g^2$. The three transition curves of relevance are: the main Gregory-LaFlamme condition requires 
\begin{equation}
	\left.\label{GLI} H^{-1/2} D V_2\right|_{horizon} < \left. H^{1/2} r_1^2 \right|_{horizon}\Rightarrow v_2<u_1^2 
\end{equation}
where we used $\varepsilon\ll 1$ as needed in the regime of interest; And the Gregory-LaFlamme condition in the T-dual geometry requires 
\begin{equation}
	\label{GLII} \frac{{\alpha'}^2}{V_2} \frac{1}{1+\frac{Q_0^2}{Q_2^2}}<r_1^2\Rightarrow \left( \frac{N_2}{N_0}\right)^2 v_2 <u_1^2\ . 
\end{equation}
Notice that in both cases, $H$ disappear! This nontrivial feature is responsible for the universal form of the black hole phase transition of interest. Note also that the transition relevant in the T-dual geometry, accessible for $S<N_0 N_2$ - a case we do not consider in this work - differs only by a factor involving the density of D0 branes per D2 brane. The T-duality condition is 
\begin{equation}
	\label{bigT} \left.H^{-1/2} D V_2 \right|_{horizon}>\alpha'\ . 
\end{equation}
where $u_0$ is the original location of the horizon in ten dimensions, $\mathcal{L}=L/\alpha'$ is the size of the smeared directions held fixed in the decoupling limit, and we have smeared $d$ transverse directions; {\em i.e.} one is considering the collapsing shell in $D=10-d$ dimensions.

Putting things together, one finds the transition curves:
\begin{itemize}
	\item For T-duality, the condition is the same as for $d=0$.
	
	\item For the first Gregory-LaFlamme condition: 
	\begin{equation}
		\label{GLIwithE} E<\frac{\mathcal{L}^d}{g^2} v_2^{\frac{7-d}{2}} . 
	\end{equation}
	
	\item For the second Gregory-LaFlamme condition in the T-dual geometry, one gets: 
	\begin{equation}
		\label{GLIIwithE} E<\frac{\mathcal{L}^d}{g^2} \left( \frac{N_2}{N_0}\right)^{5-d} v_2^{\frac{7-d}{2}}\ . 
	\end{equation}
\end{itemize}

In summary, the surviving Gregory-LaFlamme transition point in Figure~\ref{fig:simplephase} is at 
\begin{equation}
	\label{BHtransitions} E\simeq \frac{V_2^{\frac{7-d}{2}}}{G_N} 
\end{equation}
where $\sqrt{V_2}$ is the radius of the collapsing shell and $G_N=g_s^2 {\alpha'}^4/L^d$ is the gravitational constant in $D=10-d$ dimensions. This is the expected relation for the D0 black hole in $10-d$ spacetime dimensions.

\vspace{0.5in} {\bf \large Appendix B: Shell collapse} \vspace{0.5in}

In this appendix, we review the initial stages of the evolution of the shell through the traditional thin shell approximation of shell collapse. The metric outside the shell, when smeared out over $d$ directions, is 
\begin{equation}
	ds_E^2=-H^{-7/8} h dt^2+H^{1/8} \left( h^{-1} dr^2+\sum_{i=1}^d dy_i^2+r^2 d\Omega_{8-d}^2\right) 
\end{equation}
where 
\begin{equation}
	H = 1 + \frac{k}{L^d r^{7-d}}\ \ \ ,\ \ \ h = 1 - \left (\frac{r_0}{r} \right)^{7-d} 
\end{equation}
and 
\begin{equation}
	\label{keq} k= C\frac{G_{10} N}{g_s l_s} 
\end{equation}
for some numerical constant $C$.

Defining a new radial coordinate $R(\tau)$ 
\begin{equation}
	R = H^{1/16} r \\
	\Rightarrow dR = \frac{7-d + (9+d)H}{16 H^{15/16}} dr 
\end{equation}
the metric becomes 
\begin{equation}
	ds^2 = -H^{-7/8} h dt^2 + \frac{h^{-1}}{F^2} dR^2 + R^2 d \Omega_{8-d}^2 + H^{1/8} \displaystyle \sum_{i=1}^d dy_idy_i 
\end{equation}
where 
\begin{equation}
	F = \frac{7-d+(9+d)H}{16H} 
\end{equation}
The shell's world volume acquires the metric 
\begin{equation}
	ds^2 = - d \tau^2 + R( \tau)^2 d \Omega_{8-d}^2 + H^{1/8} \displaystyle \sum_{i=1}^d dy_idy_i 
\end{equation}
The extrinsic curvature tensor, on the exterior of the shell, is given by 
\begin{equation}
	K_{\theta \theta}^{+} = -R F^2 H^{7/16} \sqrt{h + \left ( \frac{\dot{R}}{F} \right )^2} 
\end{equation}
while on the interior, it has the value 
\begin{equation}
	K_{\theta \theta}^{-}= -R \sqrt{1 + \dot{R}^2} 
\end{equation}
The Israel junction conditions states 
\begin{equation}
	\gamma^i_j -\delta^i_j \, Tr \, \gamma = 8 \pi \, G_{10} \, \mathcal{S}^i_j 
\end{equation}
where 
\begin{equation}
	\gamma = K^{+} - K^{-}\ \ \ ,\ \ \ \mathcal{S} = (p+ \sigma) u \otimes u + p g 
\end{equation}

Therefore the equation of motion becomes 
\begin{equation}
	\gamma_{ij} = 8 \pi \, G_{10} \left ( g_{ij} \frac{\sigma}{8} + (p+ \sigma)u_i u_j \right) 
\end{equation}
The $\theta \theta$ component ($\theta$ being one of the angular directions on the shell) of this equation gives
\begin{equation}
	\sqrt{1 + \dot{R}^2} - F^2 H^{7/16} \sqrt{h + \left (\frac{\dot{R}}{F} \right )^2} = \pi G_{10} R \sigma \\
	= \frac{ \pi \, G_{10} \, \mu(R)}{\Omega_{8-d} L^d R^{7-d} H^{d/16}}\ , 
\end{equation}
with $\mu(R)$ being the energy of the shell in the rest frame. As the shell collapses, this thin shell approximation breaks down; as the black hole forms, the entire geometrical picture inside the hole falters. 

\vspace{0.5in} {\bf \large Appendix C: T-duality transformation} \vspace{0.5in}

In this brief Appendix, we collect the equations needed in applying T duality transformations on a 2-torus in IIA supergravity~\cite{Giveon:1994fu,Bergshoeff:1995as}. These are used repeatedly in Appendix A.

The dilaton transforms as 
\begin{equation}
	\label{Tdil} e^{2\phi''}={\alpha'}^2 \frac{e^{2\phi}}{g_{x_1 x_1} g_{x_2 x_2}+B_{x_1 x_2}^2} 
\end{equation}
The metric transforms as 
\begin{equation}
	\label{Tmetric1} {g''}_{x_2 x_2}=\frac{g_{x_1 x_1} {\alpha'}^2}{g_{x_1 x_1} g_{x_2 x_2}+B_{x_1 x_2}^2}\ \ \ ,\ \ \ {g''}_{x_1 x_1}=\frac{g_{x_2 x_2} {\alpha'}^2}{g_{x_1 x_1} g_{x_2 x_2}+B_{x_1 x_2}^2} 
\end{equation}
\begin{equation}
	\label{Tmetric2} {g''}_{ij}=g_{ij} 
\end{equation}
The NSNS B field transforms as 
\begin{equation}
	\label{TBfield} {B''}_{x_1 x_2}=-{\alpha'}^2\frac{B_{x_1 x_2}}{g_{x_1 x_1} g_{x_2 x_2}+B_{x_1 x_2}^2} 
\end{equation}
And the RR fields transform as 
\begin{equation}
	\label{TA1} {\alpha'}{A''}_t=A_{tx_1 x_2}-B_{x_1 x_2} A_t 
\end{equation}
\begin{equation}
	\label{TA3} {A''}_{tx_1 x_2}=-{\alpha'}\frac{A_t g_{x_1 x_1} g_{x_2 x_2}+B_{x_1 x_2} A_{tx_1 x_2}}{g_{x_1 x_1} g_{x_2 x_2}+B_{x_1 x_2}^2} 
\end{equation}

\providecommand{\href}[2]{#2}\begingroup\raggedright\endgroup

\end{document}